\pdfoutput=1
\documentclass[pra,showpacs,twocolumn,superscriptaddress,nofootinbib]{revtex4}

\usepackage{amsmath,amsfonts,amssymb,amstext}
\usepackage[dvipdfm]{graphicx}
\usepackage[colorlinks,linkcolor=blue]{hyperref}
\usepackage{multirow}

\newcommand{\bra}[1]{\ensuremath{\langle #1 \vert}}
\newcommand{\ket}[1]{\ensuremath{\vert #1 \rangle}}
\newcommand{\braket}[2]{\ensuremath{\langle #1 \vert #2 \rangle}}
\newcommand{\be}{\begin{equation}}
\newcommand{\ee}{\end{equation}}
\newcommand{\ben}{\begin{eqnarray}}
\newcommand{\een}{\end{eqnarray}}
\newcommand{\bF}{\begin{figure}}
\newcommand{\eF}{\end{figure}}

\newcommand{\bi}{\begin{itemize}}
\newcommand{\ei}{\end{itemize}}
\newcommand{\ud}{\mathrm{d}}
\newcommand{\mbf}[1]{\mathbf{#1}}
\newcommand{\rf}{\text{rf}}
\newcommand{\mw}{\mu\text{w}}

\newtheorem{theorem}{Theorem}
\newtheorem{lemma}{Lemma}

\begin{document}
\title{Quantum control of the hyperfine-coupled electron and nuclear spins in alkali atoms}
\date{\today}

\author{Seth T. Merkel}
\affiliation{Department of Physics and Astronomy, University of New Mexico, Albuquerque, NM, 87131, USA}

\author{Poul S. Jessen}
\affiliation{College of Optical Sciences, University of Arizona, Tucson, Arizona 85721, USA}

\author{Ivan H. Deutsch}
\affiliation{Department of Physics and Astronomy, University of New Mexico, Albuquerque, NM, 87131, USA}

\begin{abstract}
We study quantum control of the full hyperfine manifold in the ground-electronic state of alkali atoms based on applied radio frequency and microwave fields. Such interactions should allow essentially decoherence-free dynamics and the application of techniques for robust control developed for NMR spectroscopy. We establish the conditions under which the system is controllable in the sense that one can generate an arbitrary unitary on the system.  We apply this to the case of $^{133}$Cs with its $d=16$ dimensional Hilbert space of magnetic sublevels in the $6S_{1/2}$ state, and design control waveforms that generate an arbitrary target state from an initial fiducial state.  We develop a generalized Wigner function representation for this space consisting of the direct sum of two irreducible representations of $SU(2)$, allowing us to visualize these states.  The performance of different control scenarios is evaluated based on the ability to generate a high-fidelity operation in an  allotted time with the available resources.  We find good operating points commensurate with modest laboratory requirements.
\end{abstract}

\maketitle

\section{Introduction}
The control of spins is the foundation of coherent spectroscopy at the heart of NMR, atomic clocks, and many precision metrology experiments \cite{hinds08}.  More recently, spins have been seen as ideal carriers of information, with developments in spintronics for classical \cite{zutic04} and quantum \cite{awschalom02} information processing.  Atomic spin systems have been of particular interest given their excellent isolation from the environment and the available techniques in the ``quantum optics toolbox".  Examples include ensembles of atomic spins as quantum information processing elements \cite{ lukin00, polzik04, kuzmich05, kimble08, molmer08}, ion-trap quantum computers \cite{wineland04, monroe05, blatt07}, and neutral-atom optical lattices \cite{jaksch04}.  The latter has attracted tremendous attention in recent years, as controllable spin lattices are seen as a platform in which to perform quantum simulations of condensed matter systems \cite{lewenstein07} and studies of topological quantum field theory \cite{brennen07}.  

While in many studies of atomic spin control one considers two-level spin qubits, real atoms have  large spins with a rich internal structure.  The ability to fully control the Hilbert space within the atoms for various applications is an important addition to the toolbox.   It allows for the possibility of $d$-dimensional qudits as the fundamental information carriers \cite{brennen05} and the embedding of logical qubits in a qudit, which may be advantageous for control or protection from errors \cite{gottesman01}.  Additionally, manipulating a nontrivial Hilbert space allows us to explore interesting dynamics such as quantum chaos \cite{haake, ghose}.  Finally, in the same way that liquid state NMR has provided an excellent platform for exploring quantum control protocols \cite{khaneja01, vandersypen04, cory05}, atomic spin systems provide a test-bed with unique physical properties that allows for new investigations into control and measurement techniques.

In this paper we study quantum control of electron and nuclear spins of alkali atoms, coupled by the hyperfine interaction in the electronic ground state, using combinations of static, AC-radio-frequency, and AC-microwave-frequency magnetic fields.  In previous studies, we implemented similar control based on a combination of magnetic interactions and a nonlinear AC-Stark shift induced by a laser field \cite{smith04,Chaudhury2007}.  In that work, control was restricted to a single subspace of total angular momentum of the coupled spin system, rather than the whole Hilbert space.  More fundamentally, the light-shift interaction at the heart of the protocol came at the cost of some decoherence by spontaneous emission.  The maximum ratio of nonlinear light shift to photon scattering is fixed by the atomic structure, thereby limiting the ultimate utility of that approach.  In contrast, direct magnetic coupling to spins in the ground state is essentially decoherence free, with dephasing due solely to inhomogeneities and background fields that can be mitigated, in principle, by robust control techniques \cite{vandersypen04, khaneja05a}.  Rf/microwave control thus has the potential for higher-fidelity operation on a larger Hilbert space with speeds comparable to or faster than those previously achieved.  Such capabilities are similarly being explored for use in ion trap quantum information processing \cite{chiaverini07}.

Our ultimate goal is the implementation of general dynamical maps on the quantum system.  In this paper we take a first step -- preparation of an arbitrary state in the Hilbert space.  In particular, we look at open-loop state preparation through the application of control waveforms that take some particular known fiducial state to a target final state.  It was shown in~\cite{rabitz04} that this type of problem is much easier than the problem of generating arbitrary unitary evolutions, due to a mathematical property that promises local searches will find global optima.  We will discuss this in more detail in Sec.~\ref{S:stateprep}, but intuitively this problem requires specifying only one row of a $d \times d$ unitary matrix.  Creating one row requires as many free parameters as it takes to identify a single state in Hilbert space, $2 d -2$, as opposed to the $d^2 -1 $ parameters that are required to specify an arbitrary unitary matrix.

The remainder of this article is organized as follows. In Sec.~\ref{S:hamils} we establish the fundamental Hamiltonian that describes the dynamics of the spins and their interaction with the applied control fields.  In Sec.~\ref{S:controllability}, we discuss what it means to be controllable in the Lie algebraic sense \cite{jurdjevic72,Brockett73,Schirmer01} and determine which configurations of our fields satisfy these criteria.  Finally, in Sec.~\ref{S:stateprep} we present an algorithm that uses a gradient ascent method to generate control waveforms for state preparations.  We apply our protocol to the example of $^{133}$Cs, with one valence electron (spin $S=1/2$) and a nuclear spin ($I=7/2$), a total Hilbert space of dimension  $d=16$.  We show data from simulations comparing how different types of controls perform and tradeoffs that can be expected in the laboratory.  Along the way, we establish new visualization tools for coupled spin systems based on a generalized Wigner function representation and give analytic proofs for controllability of our system.  These results are discussed in the appendices. 

\section{Control Hamiltonian}\label{S:hamils}
In this work we seek to control the quantum state of a multilevel atom.  Though single atom addressing and measurement are possible \cite{schlosser01,dotsenko05,nelson07}, in practice we consider ensembles of uncorrelated particles.  To the degree that the atoms are identically prepared and uniformly addressed, with no interactions between them either from interatomic forces or through measurement backaction, we can take the joint state of the system as effectively $N$ identical copies, $\rho^{\otimes N }$.  More general many-body control is not considered here.  Restricting then to a single atom, the relevant Hilbert space of an alkali atom in its electronic ground state is the tensor product space of electronic spin $S$ and nuclear spin $I$ subsystems, $\mathcal{H}=\mathfrak{h}_S\otimes \mathfrak{h}_I$.  Given the single valence electron $S=1/2$, the Hilbert space is spanned by two irreducible subspaces of total angular momentum $F_{\pm}=I\pm1/2$, such that $\mathcal{H}=\mathfrak{h}_{+}\oplus\mathfrak{h}_{-}$.  

The Hamiltonian describing the atom and its interaction with external magnetic fields takes the form,
\be
H=A \mbf{I} \cdot \mbf{S}+2\mu_B \mbf{B}(t) \cdot \mbf {S},
\ee
where $\mu_B$ is the Bohr magneton and we have neglected the small nuclear magneton contribution.  Here and throughout we set $\hbar=1$.  We consider the application of three fields, $\mbf{B}(t)=B_0 \mbf{e}_z + \mbf{B}_{\rf}(t) + \mbf{B}_{\mw}(t)$.  The static bias field $B_0$ defines the quantization axis and Zeeman splittings between the magnetic sublevels.  The terms $\mbf{B}_{\rf}(t)$ and $\mbf{B}_{\mw}(t)$ describe magnetic fields oscillating at radio and microwave frequency, respectively.  The hyperfine coupling between spins provides an effective nonlinearity that will allow full controllability of the Hilbert space for appropriate choice of external fields.  

In the linear Zeeman regime, $\mu_B B_0 \ll A$, the static field acts separately in the two irreducible subspaces, and according to the Land\'{e} projection theorem the Hamiltonian is approximately,
\be
H_{B_0} \approx \mu_B \sum_{f=\pm} g_f \mbf{B}_0 \cdot \mbf{F} ^{(f)}\label{eq:bfield}
\ee
Here $\mbf{F}^{(\pm)} \equiv P_\pm \mbf{F}P_\pm$ refers to the total angular momentum operator projected onto the subspaces with quantum number $F_\pm$. Neglecting the nuclear magneton contribution, the g-factors for the two manifolds have equal magnitude but opposite sign, i.e. $g_+ = -g_-=1/F_+$.  The hyperfine coupling plus bias magnetic field thus determine the static Hamiltonian,
\be
H_0  = \frac{\Delta E_{HF}}{2}\left(P_+ - P_- \right)+ \Omega_0 (F^{(+)}_z - F^{(-)}_z),
\ee
where $\Delta E_{HF}=A F_{+}$ is the hyperfine splitting and $\Omega_0= \mu_B B_0/ F_+$ is the Zeeman splitting between neighboring magnetic sublevels.

\begin{figure}[b]
\begin{center}
\includegraphics[width=8cm,clip]{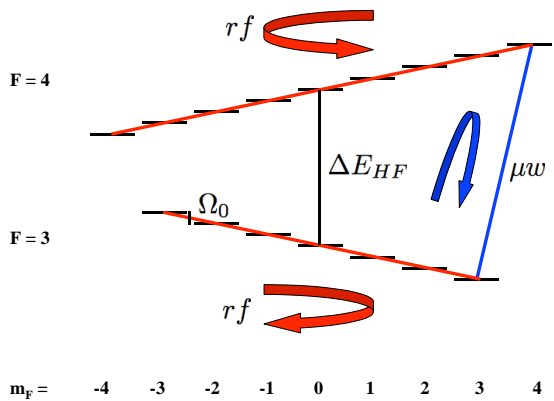}
\caption{The ground state hyperfine manifold of $^{133}$Cs.  Rf-magnetic fields (in red) lead to independent rotations in the two manifolds.  Microwaves (in blue) are the generators of rotation in a two-dimensional subspace between states in the two manifolds, here the stretched state transition $\ket{4,4} \rightarrow \ket{3,3}$. }
\label{F:hyperfine}
\end{center}
\end{figure}

As our first control field, we consider rf-magnetic fields oscillating near the frequency of the Zeeman splitting, $\omega_{\rf} \approx \Omega_0$, realized by Helmholtz coils driven with the appropriate current. We take two sets of coils that produces fields with $x$ and $y$ polarization, independent amplitude and phase control, but equal carrier frequency, $\omega_{\rf}$.  Again, for moderate current such that the amplitude of the field is in the linear Zeeman regime, the rf-Hamiltonian takes a form equivalent to the interaction with the static field
\ben
H_{\rf}(t)& =& \Omega_x (t) \cos{\big(\omega_{\rf} t - \phi_x(t)\big)}  \big(F^{(+)}_x - F^{(-)}_x\big)\nonumber\\
&+&\Omega_y (t) \cos{\big(\omega_{\rf} t - \phi_y(t)\big)}  \big(F^{(+)}_y - F^{(-)}_y\big) .
\een
The time dependent amplitudes $(\Omega_x(t), \Omega_y(t))$ and phases $(\phi_x(t),\phi_y(t))$ of the two sets of  rf coils will be used to control the system. 

To better understand the effect of the rf field, consider a resonant interaction, $\omega_{\rf}=\Omega_0$.  In the rotating frame, $H_{\rf}(t) \rightarrow H_{\rf}'(t) =U_{\rf}^{\dagger}H_{\rf}(t) U_{\rf}$, where $U_{\rf} =\exp \left\{-i \omega_{\rf}t (F^{(+)}_z - F^{(-)}_z ) \right\}$ is a rotation of the two manifolds about the $z$-axis in opposite directions, $F_x^{(\pm)} \rightarrow F_x^{(\pm)} \cos(\omega_{\rf} t) \pm  F_y^{(\pm)} \sin(\omega_{\rf} t)$, $F_y^{(\pm)} \rightarrow F_y^{(\pm)} \cos(\omega_{\rf} t) \mp  F_x^{(\pm)} \sin(\omega_{\rf} t)$.  Performing this unitary transformation and averaging over a cycle, the rf-Hamiltonian in the rotating wave approximation is,
\ben
H_{\rf}'(t) &=&\frac{\Omega_x (t)}{2} \cos{\big( \phi_x(t)\big)} \big(F^{(+)}_x -F^{(-)}_x \big)  \nonumber\\
&+& \frac{\Omega_x (t)}{2} \sin{\big( \phi_x(t)\big)} \big(F^{(+)}_y + F^{(-)}_y\big)\nonumber\\
&+&\frac{\Omega_y (t)}{2} \cos{ \big( \phi_y(t)\big)}   \big(F^{(+)}_y -F^{(-)}_y \big)\nonumber\\
 &-& \frac{\Omega_y (t)}{2} \sin{\big( \phi_y(t)\big)} \big(F^{(+)}_x+F^{(-)}_x \big). 
\label{eq:rf_hamils}
\een
Rf-control of the two spin manifolds differs from the familiar spin resonance problem.  In the latter, a single magnetic field in either the $x$ or $y$-direction would be sufficient to generate the entire $SU(2)$ algebra for rotations.   With two irreducible manifolds there is an added freedom -- the two angular momenta $F_+$ and $F_-$ can rotate in the same or opposite directions.  Amplitude and phase control of two rf-magnetic field polarizations allows us to perform arbitrary and independent rotations on the two hyperfine manifolds.  With only a single direction of $\mbf{B}_{\rf}$ we would be restricted to either co-rotating or counter-rotating in the two subspaces.  

The weak rf-magnetic fields alone will not be sufficient to fully control our atomic system; they don't couple the $F_+$ and $F_+$ manifolds, nor do they provide a nonlinear Hamiltonian within these subspaces.  In order to make our system fully controllable, we look to resonant microwaves.  While the fundamental Hamiltonian governing the microwaves is exactly of the same form as the quasistatic magnetic fields, the resonant behavior leads to very different dynamics than the previous interactions.  Depending on the polarization and frequency, the microwave couples a Zeeman sublevel in $F_+$ manifold with one in the $F_-$ manifold whose magnetic quantum number differs by $\Delta m \pm 0,1$.  For a sufficiently strong bias $B_0$ we can ignore any off-resonant excitation, and restrict the Hamiltonian to act only on a 2D subspace spanned by the states we are trying to couple.  In that case the microwave Hamiltonian has the form
\be
 H_{\mw}(t) = \Omega_{\mw}(t) \cos{\big( \omega_{\mw} t - \phi_{\mw}(t) \big)}\sigma_x,
\ee
where $\sigma_x$ is the Pauli sigma-$x$ matrix for this pseudospin, $\sigma_x = \ket{F_+,m_+}\bra{F_-,m_-} + \ket{F_-,m_-}\bra{F_+,m_+}$ and $\Omega_{\mw}(t)$ is the (time-dependent) Rabi frequency depending on the microwave power and the transition matrix element.  Again, the amplitude and phase of the microwave fields are control parameters.  In this subspace, the problem takes the form of the standard two-level resonance problem.  We must take care in going to the rotating frame to account for the simultaneous transformation we perform due to the rf-fields.  The complete frame transformation is achieved by the unitary
\be
U=U_{\rf} \exp\left\{-i \frac{\alpha t}{2} \left(P_+ -P_-\right) \right\},
\ee
where $\alpha = \omega_{\mw} -(m_+ + m_-)\omega_{\rf}$. Under this transfromation, the Hamiltonian in the rotating wave approximation for resonant microwaves is 
 \ben
 H'_{\mw}(t) &= &\frac{\Omega_{\mw}(t)}{2} \cos{\big(  \phi_{\mw}(t) \big)}\sigma_x \nonumber\\
 &+& \frac{\Omega_{\mw}(t)}{2} \sin{\big(  \phi_{\mw}(t) \big)}\sigma_y,\label{eq:mw_hamil}
 \een
generating rotations of this pseudo-spin on the Bloch sphere.

Combining the static, rf, and microwave interactions the final Hamiltonian in the rotating frame is
\be
H'(t) = H'_0+H_{\rf}'(t)+H_{\mw}'(t).
\ee
Allowing for a finite detuning of the oscillating fields from resonance, the static Hamiltonian in the rotating frame becomes, 
\be
H'_0 =  \frac{\Delta_{\mu w }}{2}\left(P_+ - P_- \right)+ \Delta_{\rf}(F^{(+)}_z - F^{(-)}_z),
\ee
where $\Delta_{\mw}=\omega_{\mw}-\Delta E_{HF} -(m+m')\omega_{\rf}$ is the effective detuning of the microwaves from the two-level transition of interest, $\ket{F_-,m_-}\rightarrow\ket{F_+,m_+}$, and $\Delta_{\rf}=\omega_{\rf}-\Omega_0$ is the rf detuning.  This, together with Eqs. (\ref{eq:rf_hamils},\ref{eq:mw_hamil}), defines the Hamiltonian we employ for control, and which we will analyze for use in arbitrary state preparation.
  
\section{Controllability}\label{S:controllability}
	In order to perform state preparation on a system, we must first determine the conditions under which the system is controllable in principle.  Is it possible to create an arbitrary state using the Hamiltonian dynamics we have available, neglecting technical constraints such as bandwidth and slew rates that restrict the types of waveforms we use to drive our atoms?  To answer this, it is simpler to analyze the conditions necessary to generate an arbitrary unitary evolution, a problem that has been studied in depth in the control theory literature,~\cite{jurdjevic72,Brockett73}, and more recently from a quantum information perspective~\cite{Schirmer01} .

\begin{figure*}[t!]
\begin{center}
\includegraphics[width=16cm,clip]{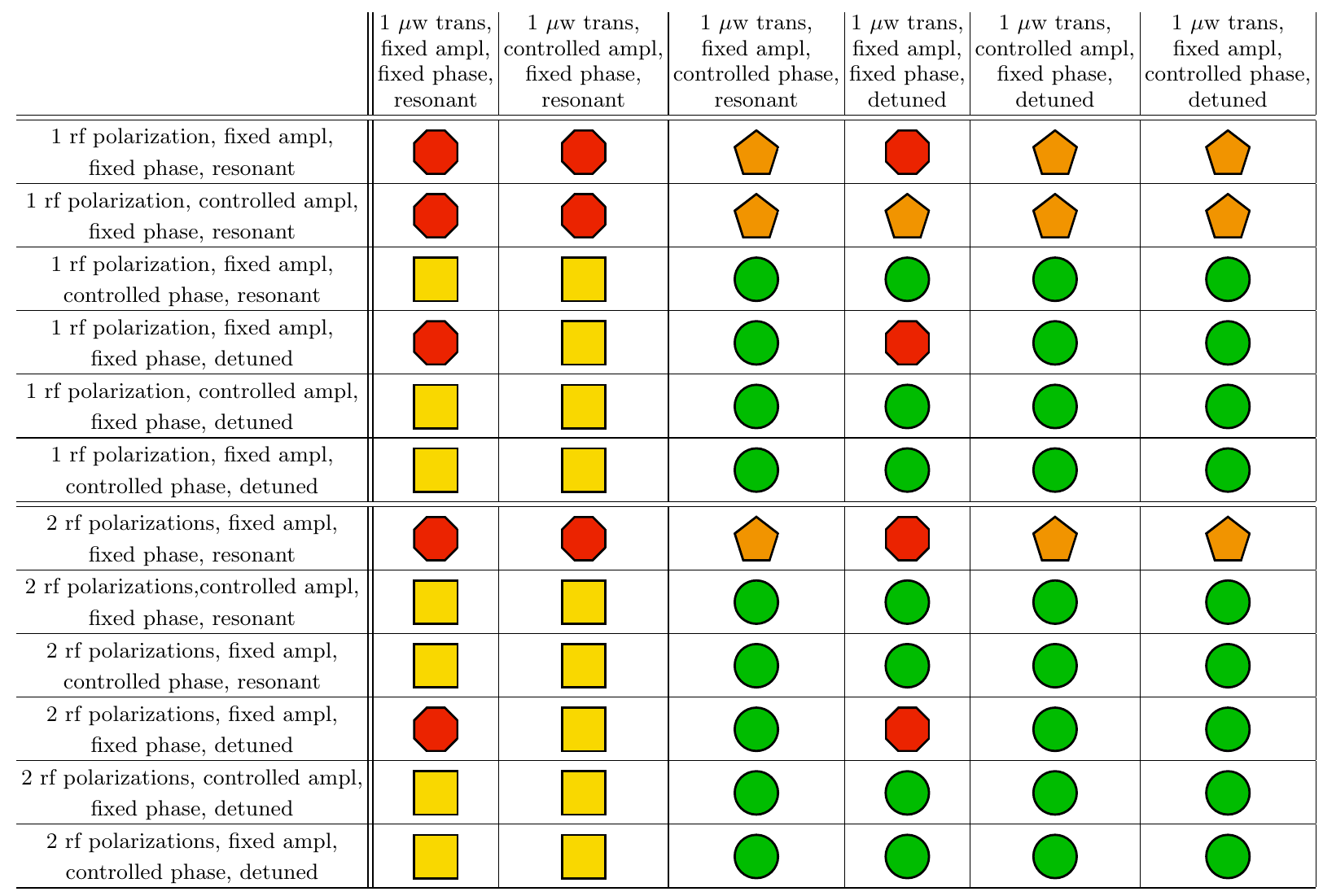}
\caption{Table exploring controllability of the system for a variety of configurations: one microwave field driven on different two-level transitions, $\ket{F=3,M} \rightarrow \ket{F=4,M'}$, amplitude and/or phase control, one or two sets of orthogonal rf coils (rf polarizations), and resonant vs. detuned fields. The different configurations yield one of four different outcomes: (green circle) all microwave transitions provide full controllability, (yellow square) all transitions but the clock transition $\ket{3,0} \rightarrow \ket{4,0}$  provide full controllability, (orange pentagon) only the transitions of the form $\ket{3,\pm 3} \rightarrow \ket{4,\pm 4}$ and $\ket{3,\pm 3} \rightarrow \ket{4,\pm 2}$  provide full controllability, and (red octagon) no transitions yield controllable Hamiltonian dynamics.  In this calculation we consider all valid microwave transitions that can be selected with polarization and/or frequency.}
\label{F:control_table}
\end{center}
\end{figure*}

	Formally stated, we consider a quantum system in a Hilbert space of dimension $d$, governed by a Hamiltonian of the form
\be
H(t) = H_0 + \sum_j b_j(t) H_j.
\ee
The system is said to be ``controllable", if for every possible unitary map, $U_0 \in SU(d)$, there exists a choice of controls $b_j(t)$ and a finite time $T$ such that the Hamiltonian evolution given by the Schr\"{o}dinger equation,~$\dot{U}(t) = -iH(t)U(t)$, maps the identity operator to $U_0$ at time $T$.  A necessary and sufficient condition for controllability is that the independent terms in the Hamiltonian $\{H_0, H_1,\ldots,H_n\}$ generate the Lie algebra $\mathfrak{su}(d)$.  Obviously, if we can perform any unitary evolution, we can perform any state preparation.  

To elucidate the connection between controllability and the generators of the Lie algebra, we review the basic principles here.  A Lie algebra is a linear vector space with an algebraic product defined by the commutator.  We can see that we can generate any linear combination of our initial set of generators by looking at very short square-pulses according to the Trotter formula, where
\be
 e^{-i H_1 \alpha \Delta} e^{-i H_2 \beta \Delta} \approx e^{-i (\alpha H_1 + \beta H_2) \Delta}.
\ee  
Such short pulses are allowed since we assume access to an arbitrary waveform.  In addition to linear combinations it is also possible to generate the commutators by 
\be
 e^{-i H_1 \Delta}e^{-i H_2 \Delta}e^{i H_1 \Delta}e^{i H_2 \Delta}\approx e^{-[H_1,H_2] \Delta^2}.
\ee     	
The ability to generate, in principle, any linear combination and any commutator means that one can simulate any element of the the Lie algebra generated by our initial independent Hamiltonians, $\{H_0, H_1,\ldots,H_n\}$, and thus any unitary contained in the associated Lie group.  If the Lie algebra generated is $\mathfrak{su}(d)$, we call this system controllable.    

For the Hamiltonian system described in Sec. \ref{S:hamils}, with arbitrary control of the amplitude and phase of the two orthogonal sets of rf-coils and a single microwave field, the control algebra generated by the six operators $\{ F_x^{(+)}, F_y^{(+)}, F_x^{(-)}, F_y^{(-)}, \sigma_x, \sigma_y \}$ is $\mathfrak{su}(d)$ in its entirety.   In this case, it is possible to prove controllability analytically for an arbitrary alkali, with an arbitrary nuclear spin $I$.  The proof is fairly involved and is shown in detail in Appendix \ref{A:a_control}.  

Though sufficient, the entire available set is not necessary to achieve controllability.  In practice, one can reduce the number of generators in the control algebra and still implement an arbitrary unitary.  For an experiment, it is important to understand which components are really necessary so that we can evaluate the tradeoffs between ease of implementation and controllability.  In order to study the capability of various reduced sets of controls we resort to numerics.  Being a linear vector space, determining whether a set of operators generates the algebra $\mathfrak{su}(d)$  only requires showing that it is possible to generate a basis for $\mathfrak{su}(d)$.  We do this by tabulating a library of all the linearly independent operators generated by the set of control operators.  

We carried out this procedure for the specific example of $^{133}$Cs with nuclear spin $I=7/2$ and study the capability of a variety of control sets to generate the entire $\mathfrak{su}(16)$ algebra.  We considered 8 different microwave configurations:  controlling or fixing the amplitude and the phase of the fields, and whether or not we are detuned from resonance.  The two cases where both the amplitude and phase are controlled and where the amplitude is fixed but the phase is controlled can be shown to be equivalent.  In the rf configurations, we also allow for one or two orthogonal sets of magnetic coils (rf polarization).  The last free parameter is the choice of which microwave transition we excite.  We assume arbitrary frequency and polarization selectivity of the desired transition for this purpose.  The results are summarized in Table~\ref{F:control_table}.  In each box we enumerate the set of microwave transitions that yield controllable dynamics.   We find that our system is controllable for a wide number of configurations, though there are some specific cases in which it is not.  For example, out of all the choices for microwave transitions, the clock transition, $\ket{F_+,0} \rightarrow \ket{F_-,0}$, renders the system controllable in the least number of scenarios.  This should not come as much of a surprise since we are controlling the system with rf magnetic fields and this transition is insensitive to magnetic fields.   

It is interesting to note that there exist configurations that are controllable in which there is one time-dependent control field and some fixed time-independent fields.  This is the simplest scenario one could expect to find, and allows for bang-bang control, a well-studied protocol.  In this paper, however, we look at the control systems that utilize more parameters, decreasing the time needed for state preparation.         

\section{State Preparation}\label{S:stateprep}
We seek to design Hamiltonian evolutions that take an initial known quantum state to an arbitrary quantum state in the Hilbert space.  Historically, numerical searches for control waveforms have performed much better than expected.  The fidelity of a state preparation is a functional of the control waveform,
\ben
F[\mbf{b}(t)]  &=& |\bra{\psi_{target}}U[\mbf{b}(t)] \ket{\psi_0} |^2\nonumber\\
&=&|\bra{\psi_{target}} \mathcal{T} ( e^{-i \int_0^T H_0 + \sum_j b_j(t) H_j \ud t } ) \ket{\psi_0} |^2,\nonumber\\
\een
where $\mathcal{T}$ is the time ordering operator.   Under ideal conditions, assuming no decoherence and an arbitrarily amount of time to perform the control, Rabitz {\em et al.}~\cite{rabitz04} proved that the control landscape is surprisingly simple -- every local optimum is a global optimum.  This implies that, 
\be 
\nabla_{\mbf{b}(t)} F[\mbf{b}_0 (t)]  = 0 \Leftrightarrow F[\mbf{b}_0 (t)] = \{ 0, 1\},
\ee
completely independent of the initial and target states.  Therefore, a local search of the space of control fields, starting from any random initial guess, will find a global maximum of the fidelity.  For this problem, gradient searches perform about as well as more computationally intensive searches like genetic or simulated annealing algorithms.  

In a real system the assumptions of the proof will not hold.  There will always be some decoherence and one does not have infinite time to perform the control.  In fact, we would like to perform state preparation as fast as possible in order to combat decoherence and various inhomogeneities that lead to accumulated errors.  Additionally, we need to consider control fields that have a limited bandwidth and slew rate constraints.  For these realistic conditions, not every gradient search from an arbitrary starting point yields a global maxima.  Nonetheless, we have found empirically that the results of the theorem are approximately true with moderate decoherence and after a sufficient time.  We still find excellent protocols after making only a small handful of searches, and these can be further filtered to find control waveforms that perform well under realistic operating conditions.

\begin{figure*}
\begin{center}
\includegraphics[width=14cm,clip]{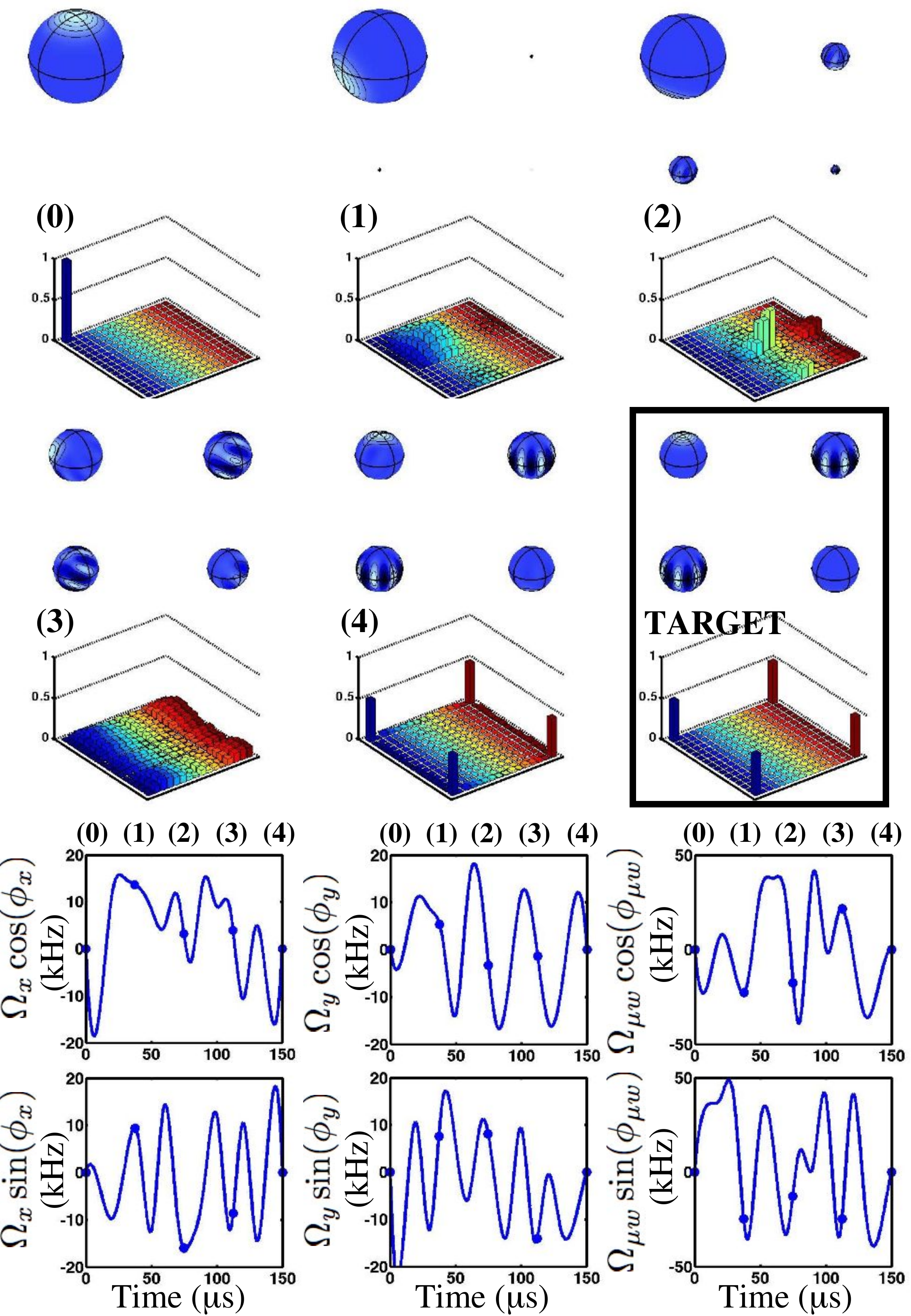}
\caption{Sample evolutions that result from our state preparation algorithm.  Starting in the spin coherent state $\ket{4,4}$, we simulate preparation of the state $\frac{1}{\sqrt{2}}(\ket{4,4}+\ket{3,-3})$ and obtain a fidelity  0.993.  We control the amplitudes and phases of rf coils in both the $x$ and $y$ directions, as well as the amplitude and phase of a resonant microwave that couples the states $\ket{4,-4}$ and $\ket{3,-3}$.  We show the Cartesian components of the three control fields ($\Omega \cos \phi$ and $\Omega \sin \phi$) over the entire state preparation time of 150$\mu$s.   We show snapshots of the evolved state at five different times, identified as times (0)-(4).  Two different representations of the state are shown: bar charts of the absolute values of the density matrix elements, and a generalized spherical Wigner function.  The spheres on the diagonal represent the Wigner functions in the irreducible subspaces $F^{\pm}$ and the off-diagonal spheres represent the coherences between the manifolds.  For details see Appendix B.}
\label{F:time_series1}
\end{center}
\end{figure*}

\begin{figure*}
\begin{center}
\includegraphics[width=14cm,clip]{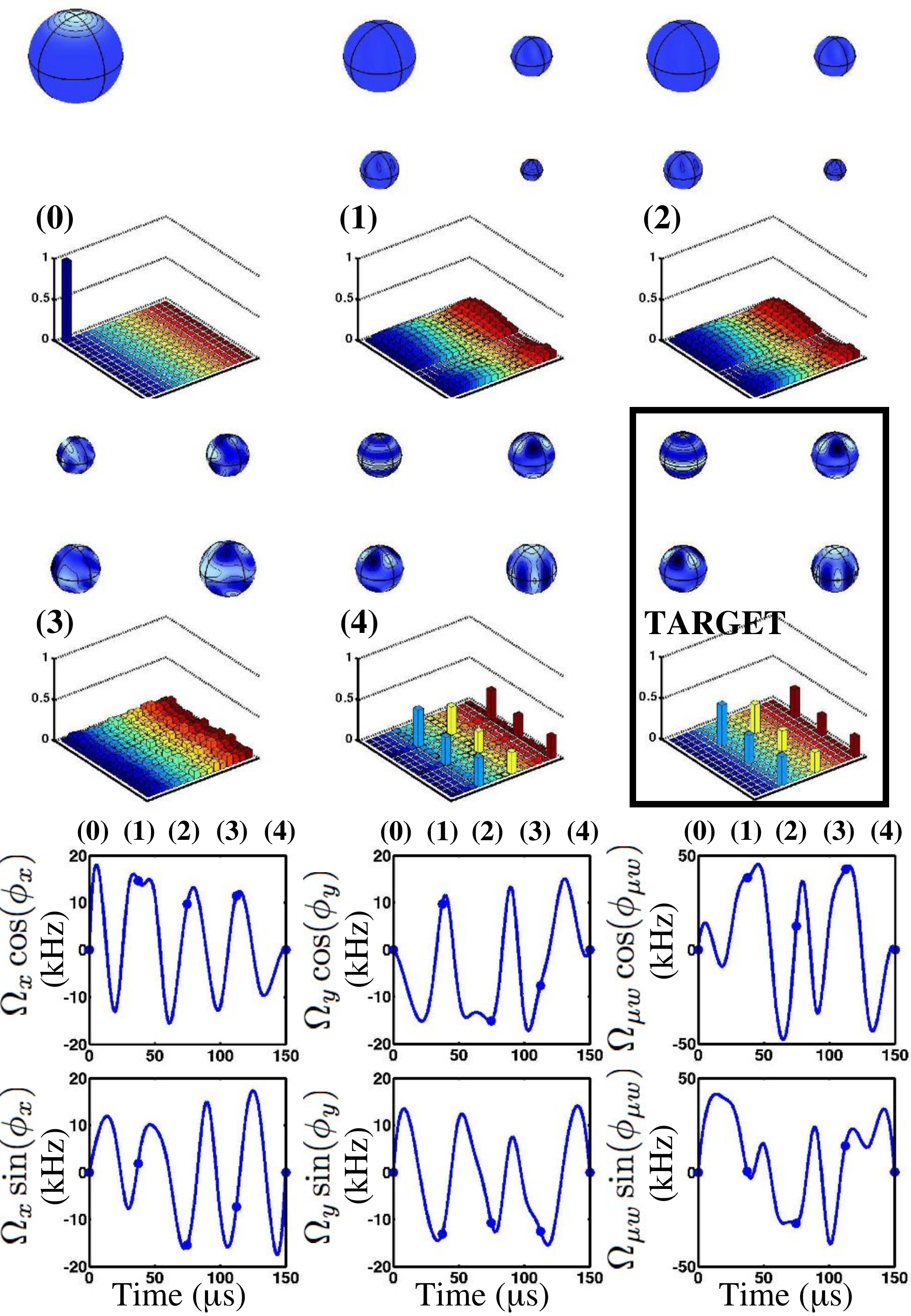}
\caption{Same as Fig.~(\ref{F:time_series1}), preparing the state  $\frac{1}{\sqrt{2}}\ket{4,4}+\frac{1}{2}(\ket{3,3}+\ket{3,-3})$ with a fidelity of 0.995.}
\label{F:time_series2}
\end{center}
\end{figure*}

\subsection{Optimization protocol}
As we are dealing with the optimization of waveforms that are functions of continuous time, the first step is to transform the problem into a search for a finite number of values at discrete times.  The physical constraints of bandwidths and slew rates of the controllers provide a natural scale. There is a minimum interval during which a field can vary over a maximum range.  A discretized version of a control waveform is thus specified as a vector of values within this range at these fixed intervals.  The continuous control waveforms are then found by interpolation using cubic splines, consistent with the bandwidth constraints, at least on a fine enough grid for use in our numerical integration of the Schr\"{o}dinger equation. 
  
We create optimal control waveforms by first fixing the total time of the state preparation procedure.  Due to our discretization technique, fixing the total time fixes the number of optimization variables.  Starting from a randomly chosen initial vector of control waveform values, $\mbf{b}_0$, we perform a gradient ascent search by taking small steps in the direction of steepest ascent, i.e. 
\be
\mbf{b}_{n+1} = \mbf{b}_n + \epsilon \nabla F(\mbf{b}_n).
\ee   
An optimal value corresponds to the maximum, where the gradient approaches zero.  We performed this search numerically on a Matlab cluster by optimizing waveforms from a handful of random seeds in parallel, and then chose the one that gave the highest fidelity.  In this work we do not consider the robustness of the waveform to inhomogeneities and noise.  More complex objective functions can be optimized as in~\cite{Chaudhury2007} once the relevant experimental conditions are known.

We applied this protocol to the specific case of $^{133}$Cs, with ground-state hyperfine splitting of $\Delta E_{HF} = 9.2$ GHz.  We take a static bias field to produce a Zeeman splitting of $\Omega_0= 1.0$ MHz, sufficient to give excellent resolution of the magnetic sublevels, but well within the linear Zeeman regime.  The rf field power is chosen so that on resonance the rotation rate is characterized by $\Omega_{\rf} = 15$ kHz.  As a generic case, we take one microwave field, resonant on one of the stretched transitions $\ket{F=3,M=\pm -3} \rightarrow \ket{F=4,M= \pm -4}$, where the microwave Rabi frequency is largest, and the system is controllable in a wide variety of scenarios.  The microwave power is chosen to give a Rabi frequency $\Omega_{\mw} = 40$ kHz.  The slew rates constrain the maximum rate of change of amplitude and phase of the control fields.  In the case of the rf-magnetic field, a ``slew time" of $\tau_{\rf} =10 \mu$s fixes the slew rates on the amplitude to 1.5 kHz$/\mu$s and phase to 0.2 $\pi/\mu$s.  In the case of microwaves, faster control is possible, with a slew time of $\tau_{\mw} = 1.0$ $\mu$s, or amplitude and phase slew rates of 40 kHz$/\mu$s and 2.0 $\pi/\mu$s respectively.

Two examples of the end product of this optimization are shown in Figs.~(\ref{F:time_series1},\ref{F:time_series2}) for target states $\frac{1}{\sqrt{2}}(\ket{4,4}+\ket{3,-3})$ and $\frac{1}{\sqrt{2}}\ket{4,4}+\frac{1}{2}(\ket{3,3}+\ket{3,-3})$ respectively.  The initial state for these examples is the stretched state $\ket{4,4}$, a state easily reached by optical pumping.  We optimized waveforms for both the amplitude and phase for two rf polarizations in the $x$ and $y$ directions as well as for the microwave field. The fidelities of preparation in both cases are greater than 99\%.  The state preparations shown here take 150$\mu$s, an interval that ensures that a moderate search will yield high-fidelity waveforms.  More intensive optimizations can yield faster control waveforms.  

Our gradient search algorithm leads to waveforms that cause the system to undergo quite complex dynamics, as evidenced by the intermediate states seen in the course of the evolutions,  Figs.~(\ref{F:time_series1},\ref{F:time_series2}).  One may wonder whether there are simpler choices, since given a fixed initial state, there are many different waveforms that lead to same target state.  While our method does lead to waveforms that are hard to intuitively understand, some recent studies \cite{schirmer08} suggest that the waveforms derived from gradient searches may be more robust than those that come from more geometric algorithms.

\subsection{Performance of Optimization} 
\begin{figure*}[t!]
\begin{center}
\includegraphics[width=16cm,clip]{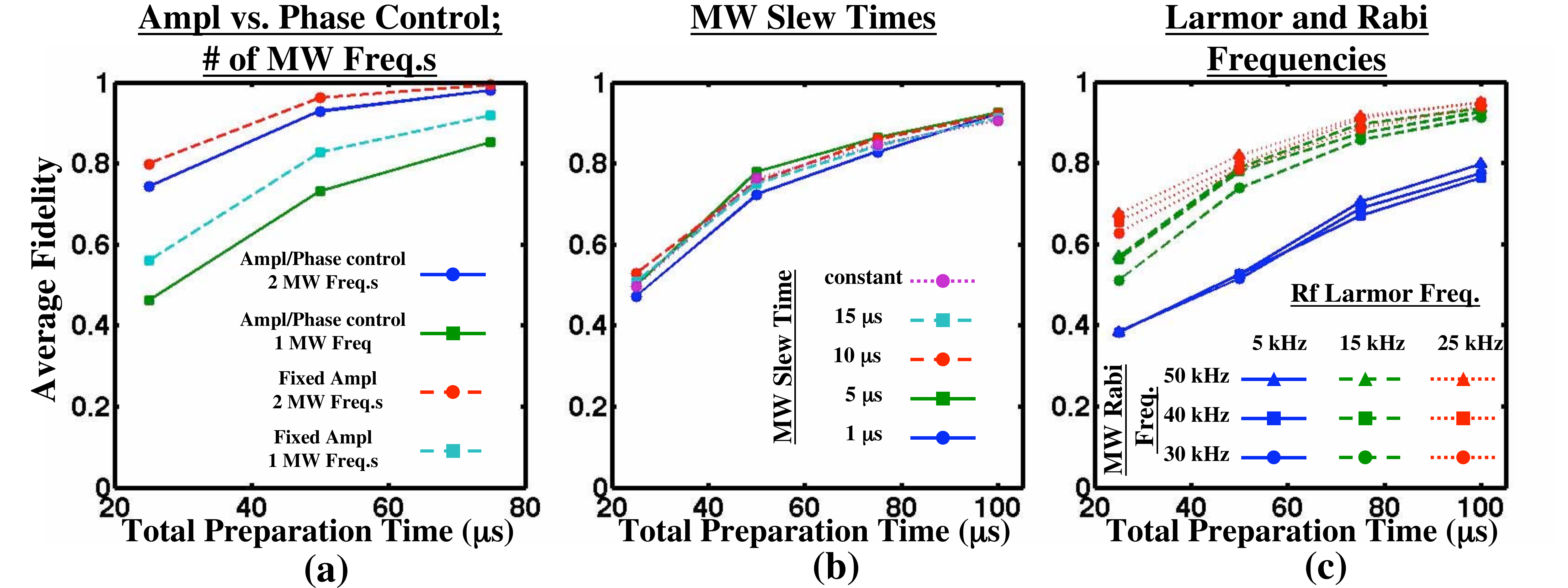}
\caption{Plots of the average fidelity of state preparation for different control configurations and total preparation times.  Each point represents the fidelity averaged over a set of 10 states randomly chosen from the Harr measure.  For each state and configuration, the gradient search was performed with 20 random seeds and we chose the protocol that generated the highest fidelity.}
\label{F:opt_stats}
\end{center}
\end{figure*}

In Sec.~\ref{S:controllability} we discussed the mathematical conditions necessary for our Hamiltonian dynamics to be controllable.  These conditions, while useful for ruling out large classes of Hamiltonians as unsuitable for our purposes, tell us nothing about the relative performance of different control scenarios.  Our figure of merit is the time after which we can be reasonably sure that our optimization will find a high fidelity waveform for any target state.  To determine this time for a given control protocol, we run our optimization up to a given final control time over a large collection of randomly chosen states and determine the average fidelity.  In this section we examine these results and discuss some of the tradeoffs and bottlenecks that might be encountered in the lab.

There are many parameters in this system that we can manipulate, including the number of independently controlled rf polarizations, the number of resonant microwave frequencies, the types of controls (amplitude vs. phase), detuning, slew rates, and the strengths of the different fields.  Based on some of our previous experiments we set as a baseline: one microwave frequency, two orthogonal rf polarizations,  rf power giving $\Omega_{\rf}$ = 15 kHz, a microwave Rabi frequency of $\Omega_{\mw}$ = 40 kHz, a rf slew time of 10 $\mu$s, and a microwave slew time of 1.0 $\mu$s.  While we could independently vary all these parameters, this would be an unwieldy computation.  Here we fix some of the parameters that are unlikely to differ in the future experiments we are considering.  In particular, we fix the rf slew time to be 10 $\mu$s and consider control with two sets of rf coils.  For simplicity we also consider all fields to be resonant, and the microwaves to couple the stretched states.        

Statistics were collected by running the state preparation algorithm for 10 different random states found by sampling using the Harr measure on $SU(16)$ \cite{Pozniak98}.  In all cases the initial state was the $\ket{4,4}$ state.  For each combination of total time, target state, and system configuration, we run the optimization 20 times starting from different random seeds of the vector that defines the control waveform, as discussed is Sec.~\ref{S:controllability}.  Out of this set of 20, we choose the highest fidelity preparation.  The fidelities from the 10 random states are averaged to produce the data points shown in Fig.~\ref{F:opt_stats}.  In principle, more iterations would yield higher fidelity waveforms, but it is useful to understand which types of high-fidelity controls can be found after only modest searches.  

In Fig.~\ref{F:opt_stats}a, we study the effect of varying the characteristics of the microwave field. We compare the performance of one vs. two resonant microwave frequencies on one or both of the stretched transitions, $\ket{3,3} \rightarrow \ket{4,4}$ and  $\ket{3,-3} \rightarrow \ket{4,-4}$.  In addition, we examine the effect of removing control of the microwave amplitude (a scenario that still allows for full controllability of the system, as discussed in Sec.~\ref{S:controllability}).  As expected, since the microwave Rabi frequency is larger than the rf Larmor frequency, increasing the number of microwave fields has a large effect.  On the other hand, it was surprising that fixing the microwave amplitude, thereby substantially decreasing the number of control parameters, yielded higher fidelity waveforms.  We suspect that while there most likely exist higher fidelity waveforms with control of both amplitude and phase, increasing the number of microwave control parameters rapidly increases the dimension of the search space, requiring many more iterations of our algorithm to find a superior waveforms, on average. This suspicion is reinforced by Fig.~\ref{F:opt_stats}b, where we consider the effect of microwave slew time.  With our baseline parameters, it would appear that increasing the microwave slew time doesn't really limit the optimized control performance.  In fact, the smallest slew time we considered, 1.0 $\mu$s, performed slightly worse than the other slew times, including the case where the microwave amplitudes are fixed.  As a reminder, the slew time determines the information content of our waveforms, and thus the number of optimization variables.  Again, we see that for the modest searches we are performing, decreasing the dimension of the search space counterbalances the loss of control.

In Fig.~\ref{F:opt_stats}c, we study the effect of the power in the rf and microwave fields.  For these simulations we fixed the amplitudes of the fields and solely control their phases.  We find that varying the microwave power around our baseline makes little difference.  The rf power is slightly more important, but increasing the Larmor frequency above the baseline has a fairly  small effect.  These results indicate that the slew rate and bandwidth constraints we have imposed on the rf magnetic fields are the bottleneck for controlling the system, and limit the ability to more rapidly control the system through increases in power.  It would appear that the microwave parameters we employ as our baseline are also well above the limits imposed by this bottleneck and we can safely reduce the microwave power and slew rates without sacrificing performance. The rf Larmor frequency we employ is commensurate with the slew rate constraint.           

By optimizing many state preparations for a variety of control configurations we find state preparation protocols with this system that take between $50-150 \mu$s.  We can compare this to the types of control waveforms that were implemented in our previous work that employed a nonlinear AC-Stark shift to achieve controllability~\cite{Chaudhury2007}.  The waveforms we find here are about an order of magnitude faster, control a Hilbert space that is double the dimension, and have negligible decoherence as compared to the intrinsic decoherence that arises from spontaneous emission.

\section{Summary and outlook}
In this paper we have studied quantum control of the $d=2(2 I+1)$ dimensional Hilbert space associated with coupled electron spin $S=1/2$ and nuclear spin $I$ of alkali atoms in their electronic ground state, based on interactions with static, rf, and microwave magnetic fields.  Such interactions allow rapid and essentially decoherence-free dynamics.  We studied a variety of configurations that allow for full controllability of the system based on analytic proofs for the most general control fields considered and numerical studies in more restricted configurations.  With controllability in hand, we studied the problem of open-loop state preparation, mapping a known fiducial state to an arbitrary target state of the Hilbert space, applied to the specific problem of $^{133}$Cs with a $d=16$ ground-electronic subspace.  Control waveforms can be found from simple gradient searches in the control landscape.  We evaluated the performance of a variety of scenarios, restricting some control parameters by e.g. fixing the amplitudes of the fields or the number of resonant microwaves frequencies.  We find that under certain conditions, restricted control yielded better performance.  We attribute this to the complexity in searching a large dimensional control parameter space.  

Implementation of the proposed control protocol discussed in this article will require diagnostics to measure the fidelity of the prepared state with respect to the target.  This can be done via quantum state tomography on the ensemble.  In prior work, we developed and implemented a protocol whereby the quantum state is estimated via continuous measurement on a single ensemble of identically prepared atoms \cite{silberfarb05,smith06}.  To achieve this, the system must be controllable. By applying a well-chosen waveform in the course of the  continuous measurement, one gains access to an informationally complete record. One then  inverts the measurement history to obtain a high-fidelity estimate of the initial state, limited only by the signal-to-noise ratio and decoherence that occurs during the measurement.  In our prior work, combinations of laser interactions and magnetic fields were employed to yield an informationally complete measurement record.  In that case, the laser beam acted both as a probe of the atoms to provide the measurement record and as a control field to provide a nonlinear light-shift on the atoms.  To extend this protocol to the case at hand, we must control the system to produce an informationally complete set of observables on the full $d$-dimensional Hilbert space.  Using the microwave/rf  interactions studied here, one can achieve this while separating the control and measurement functions of the applied fields. This should be faster and reduce decoherence induced by photon scattering of the probe.  We will study this in future work with the goal of rapid and robust quantum state estimation in a large dimensional Hilbert space.  

In addition to state preparation, Lie-algebraic controllability implies that there exist waveforms to generate arbitrary unitary maps on the system.  The work considered here corresponds to implementing one column of a unitarity because the waveforms we design lead to the intended dynamics only on a single fiducial quantum state.  Nonetheless, such capabilities provide a starting point to more general control tasks such as embedding a qubit in a qudit, or the implementation of universal qudit control. Moreover, because the Hamiltonians considered here correspond to generators of rotations either within irreducible subspaces or on pseudospins, it gives us a natural starting point to consider the application of composite pulse sequences developed for NMR \cite{vandersypen04} in order to make our protocols more robust to the inevitable imperfections in our system. 

This research was supported by NSF Grants No. PHY-0653599 and No. PHY-0653631, ONR Grant No. N00014-05-1-420, and IARPA Grant No. DAAD19-13-R-0011.

\appendix
\section{Analytic Controllability}\label{A:a_control}
We prove here that our quantum system is controllable given accesses to the Hamiltonian presented in Sec.~\ref{S:hamils}.  More precisely we show that the set of operators $\{F^{(+)}_x,F^{(+)}_y,F^{(-)}_x,F^{(-)}_y, \sigma_x, \sigma_y \}$ generates the Lie algebra $\mathfrak{su}(d)$, where $d$ is the dimension of the tensor product space, $d=2(2I+1)$.   In the language of control theory, a set of operators is said to ``simulate" another operator if we can construct this operator through linear combination and commutators.  Starting with a generating set, one simulates new operators which are added to our library.  The goal is to use the generators to simulate a basis for the entire Lie algebra.

We begin by proving a very general theorem relating to controllability when one is able to perform SU(2) rotations on an n-dimensional Hilbert space.

\begin{theorem}\label{t:rank_2}
In an $d$-dimensional Hilbert space with $d>2$, if one has access to the irreducible generators of rotations, $J_x$ and $J_y$, then in order to fully control the space it is sufficient to add an operator $h$ that has a non-zero overlap (according to the trace inner product) with at least one rank-2 irreducible spherical tensor.  That is 
\be
 \exists ~q~ \textrm{s.t.} ~Tr\left( h T^{(2)}_q \right) \neq 0\quad \Rightarrow \quad\{J_x, J_y, h\}_{L.A.} = \mathfrak{su}(d).\nonumber
 \ee
\end{theorem}

Here we have introduced the orthonormal basis of irreducible spherical tensor operators, 
\be 
\label{irreducible}
T^{(k)}_q(J) = \sqrt{\frac{2k+1}{2J+1}}\sum_m \braket{J,m+q}{k,q;J,m} \ket{J,m+q}\bra{J,m},
\ee
satisfying the fundamental commutation rules,  
\ben
\left[J_z, T^{(k)}_q \right] &=& q T^{(k)}_q \\
\left[J_\pm, T^{(k)}_q \right] &=& \sqrt{k(k+1)-q(q\pm1)}T^{(k)}_{q \pm 1} \nonumber,
\een
where $J_\pm = J_x \pm i J_y$.   It follows from these commutators that given the set $\{J_x, J_y, T^{(k)}_q\}$ one can simulate any rank-$k$ irreducible tensor, and since these are an operator basis, the generators of rotation can map any rank-$k$ operator to any other rank-$k$ operator.  With this property we are now prepared to prove a lemma.

\begin{lemma}\label{t:Z2}
 $\{J_x, J_y, T^{(2)}_0\}$ generates $\mathfrak{su}(d)$.
\end{lemma}
We prove this by first noting that
\be
\left[ T^{(2)}_0, T^{(k)}_q\right] = c_{k,q} T^{(k+1)}_q + d_{k,q} T^{(k-1)}_q.
\ee
The exact form of the constants is irrelevant except for the fact that there is always some rank-$k$ tensor for which $c_{k,q}$ is nonzero.  Given this, the proof follows by induction.  Suppose our library of simulatable operators contains all operators of ranks $k$ and $k-1$. By commuting some rank $k$ operator with $T^{(2)}_0$ we obtain an operator with support on operators of rank $k-1$ and $k+1$, thus containing a component in the space of rank $k+1$ operators that is linearly independent from the current set of Hamiltonians in our library.  Commutation with the generators of rotation allow us to simulate all other rank $k+1$ operators.  Since we can simulate all rank-1 from the generators $\{J_x,J_y\}$, and the rank-0 operator is the trivial identity operator, it follows by induction that we can simulate all rank-$k$ operators that are supported on the Hilbert space, $k \le d-1$.  Therefore $\{J_x, J_y, T^{(2)}_0\}$ generates $\mathfrak{su}(d)$.QED   

With this lemma, we see that in order to show theorem \ref{t:rank_2}, we need merely to show that the set $\{J_x, J_y, h\}$ can simulate the operator $T^{(2)}_0$.  We will do this in essentially three steps.  Before we start we expand the Hamiltonian $h$ in our spherical basis, $h = \sum_{k=1}^{d-1} \sum_{q=-k}^{k}  h^{(k)}_q T^{(k)}_q $.

\subsection{Step 1: Simulate $h_1 = T^{(2)}_0 +  \sum_{k=3}^{d-1}  \sum_{q=-k}^{k}  h'^{(k)}_q T^{(k)}_q$ }

To simulate $h_1$ we note that $h$ is defined to have some nonzero rank-2 component.  With rotations we can transform the rank-2 component to $T^{(2)}_0$.  Additionally, since we have all the rank-1 tensors in our library already, we can remove the rank-1 piece of $h$ through linear combinations to yield $h_1$.
   
\subsection{Step 2: Simulate $h_2 = T^{(2)}_0 +  \sum_{k=3}^{d-1}  h''^{(k)}_0 T^{(k)}_0 $ }

Consider the double commutator 
\be
\left[ J_z, \left[ J_z,h_1\right] \right]=\sum_{k=3}^{d-1}  \sum_{q=-k}^{k} q^2 h'^{(k)}_q T^{(k)}_q.
\ee
If we take a linear combination $h_1 -a \left[ J_z, \left[ J_z,h_1\right] \right] $ the resulting operator has the same coefficients for $q=0$. For $q_0 \neq 0$, choosing $a=1/q^2$, we can sequentially remove all rank-2 tensor components, and we are left with $h_2$.

\subsection{Step 3: Simulate $T^{(2)}_0$ }

Consider the double commutator
\ben
[J_x,[J_x,h_2]] = && \frac{3}{2}T^{(2)}_0+\frac{\sqrt{6}}{2}(T^{(k)}_2+T^{(k)}_{-2})  \nonumber\\
&+&\frac{1}{4} \sum_{k=3}^{d-1} h''^{(k)}_0  \bigg( 2 k (k+1) T^{(k)}_0  \nonumber\\
&+&\sqrt{ (k-1) k (k+1) (k+2)}(T^{(k)}_2 + T^{(k)}_{-2}) \bigg). \nonumber \\
\een

We repeat the process in Step 2 to remove the components from $h_2$ with $q \neq 0$ to obtain
\be
h_2' =  \frac{3}{2} T^{(2)}_0  + \sum_{k=3}^{d-1}  a''^{(k)} \frac{ k (k+1)}{2} T^{(k)}_0.
\ee
If we now take the linear combination $h_2 - 2 h_2' /(k_0(k_0+1))$ we remove the $T^{(k_0)}_0$ component, but are left with a nonzero $T^{(2)}_0$ term.  Repeating this procedure for $k_0 = 3\ldots (d-1)$ yields an operator that is proportional to $T^{(2)}_0$.  This completes our proof of theorem \ref{t:rank_2}. 

With theorem \ref{t:rank_2} in hand we are now prepared to show that  $\{F^{(+)}_x,F^{(+)}_y,F^{(-)}_x,F^{(-)}_y, \sigma_x, \sigma_y \}$ generates the Lie algebra $\mathfrak{su}(d)$.  Unfortunately, we do not start off with the ability to perform the irreducible generators of rotations on the entire space, and so cannot immediately prove controllability using theorem \ref{t:rank_2}.  We do, however, have access to the generators of angular momentum for both the $F_+$ and $F_-$ subspaces, and so we begin by showing that we can simulate any operator that has support on only one of the two manifolds. 

To show controllability of the $F_+$ manifold we require an operator that has a nonzero overlap with a rank-2 tensor on that space.  Restricted to the $F_+$ subspace, the $\sigma_z$ operator looks like a projector onto some particular sublevel, $\ket{F_+,m_+}\bra{F+,m_+}$.  The overlap of this projector with $T^{(2)}_0$ is $\textrm{Tr}\left( \ket{F_+,m_+}\bra{F_+,m_+} T^{(2)}_0\right) = \sqrt{5/11} \braket{F_+,m_+}{2,0;F_+,m_+}$, which is nonzero for all values of $m_+$.  Of course, $\sigma_z$ has support in the $F_-$ manifold.  However, $\left[ F^{(+)}_x,\sigma_z\right]$ is confined to the $F_+$ manifold.  Since commuting by $F^{(+)}_x$ can't change the rank of a tensor, we are left with an operator confined to the $F_+$ manifold that has a nonzero overlap with some rank-2 tensor, and so according to theorem \ref{t:rank_2}, we have complete control of the $F_+$ manifold.  This proof directly carries over to the $F_-$ manifold.

\begin{figure*}[t!]
\begin{center}
\includegraphics[width=15.15cm,clip]{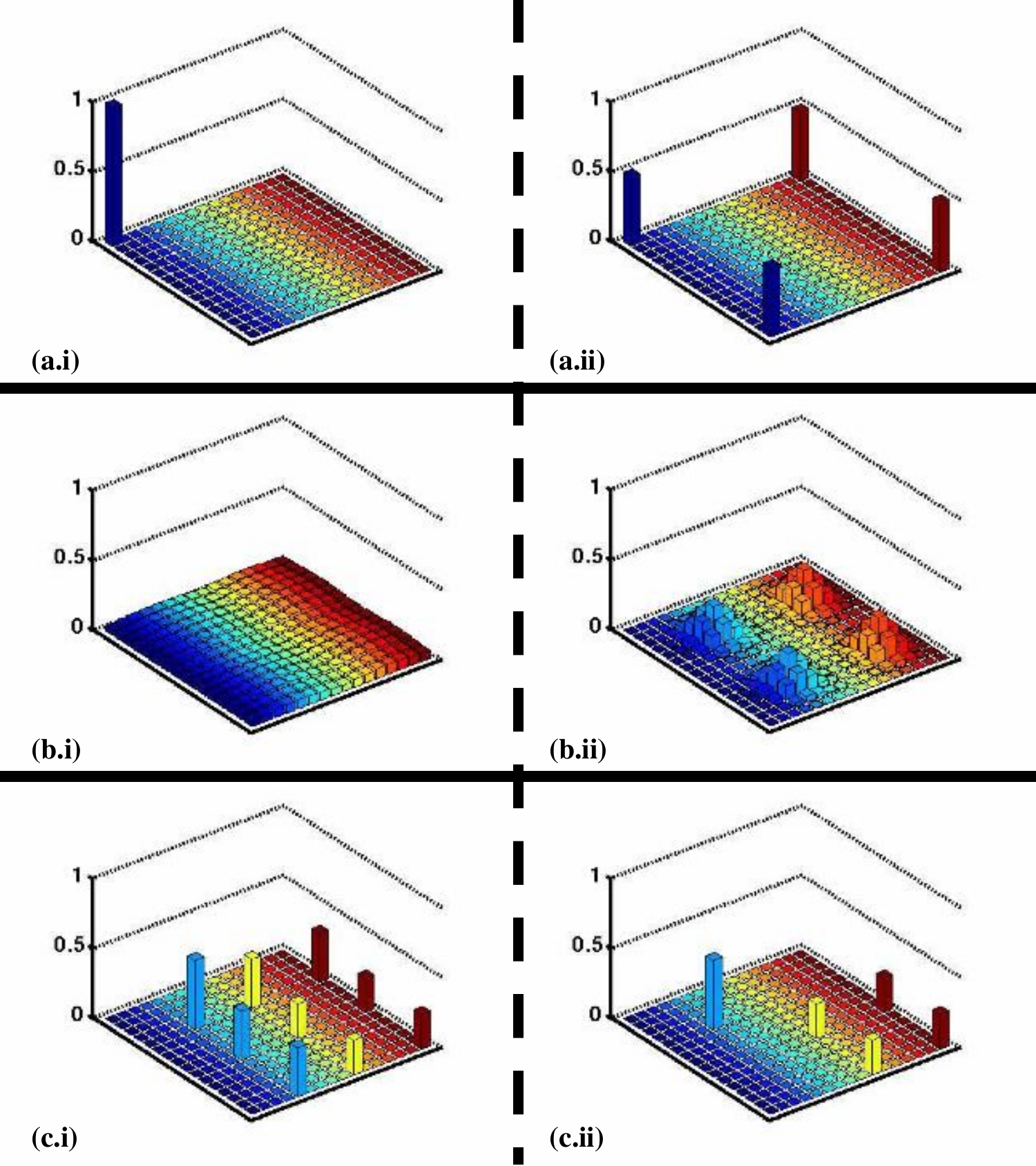}
\caption{Representations of states with bar charts of the absolute values of the density matrix elements.  (a.i) is a spin coherent state $\ket{\psi}_{ai} =\ket{4,4}$ and (a.ii) is a superposition two oppositely oriented spin coherent states, one for each of the two manifolds,    $\ket{\psi}_{aii} =\frac{1}{\sqrt{2}}(\ket{4,4}+ \ket{3,-3}$.  In (b.i, b.ii) we show the effects of rotations on a superposition of spin squeezed states, each determined as the ground state of $F_z^2-F_y$ in the respective irreducible manifold. Finally, in (c.i) we have a coherent superposition of the state $\ket{4,0}$ and a cat state $\frac{1}{\sqrt{2}}(\ket{3,3} + \ket{3,-3})$ and in (c.ii) we have an incoherent mixture of those two states. }
\label{F:plots1}
\end{center}
\end{figure*}

\begin{figure*}[t!]
\label{plots2}
\begin{center}
\includegraphics[width=15.15cm,clip]{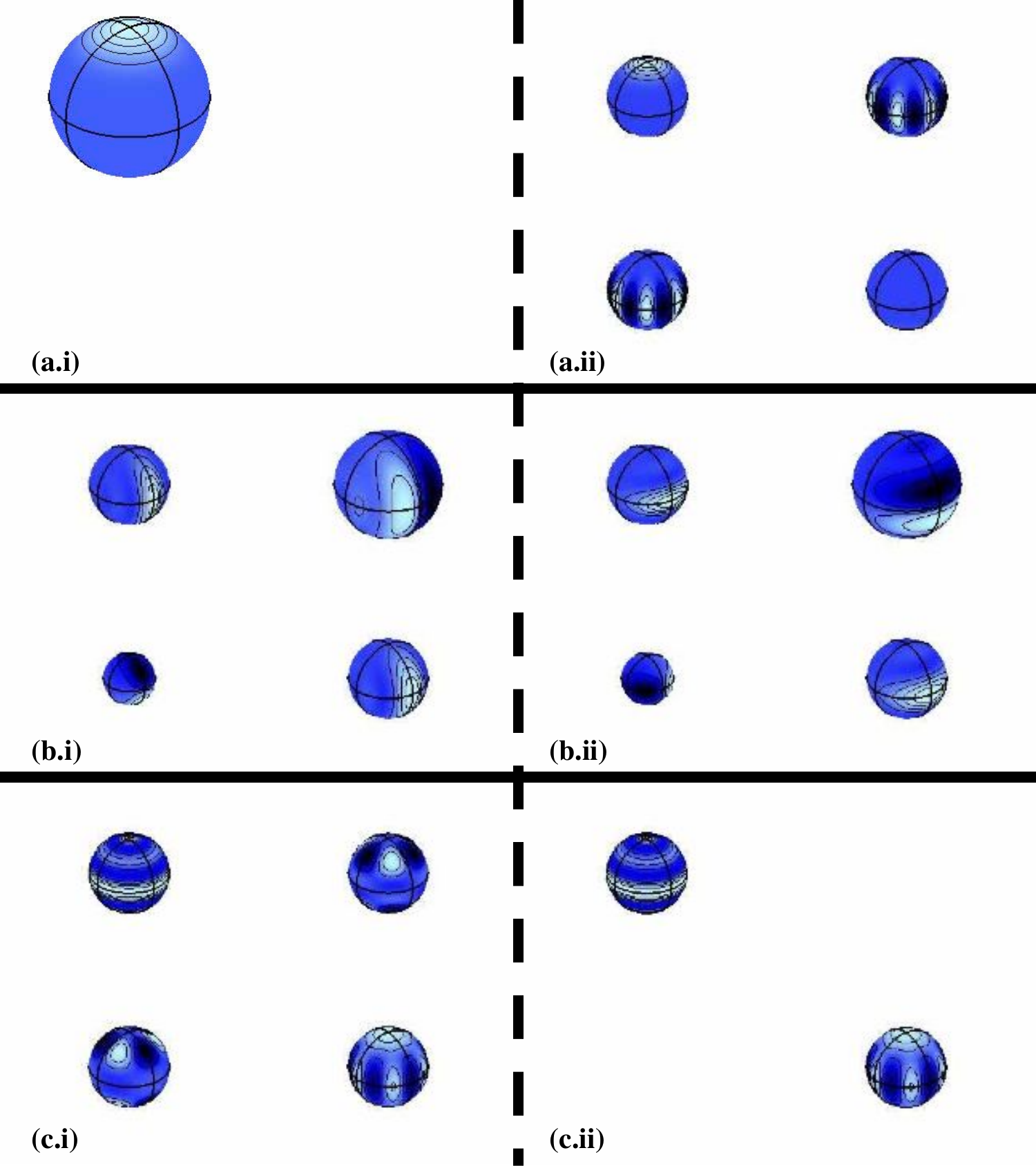}
\caption{Representations of the six states shown in Fig.~(\ref{F:plots1}) by the generalized spherical Wigner functions. Each state is represented by four spheres.  The spheres on the diagonal are the standard SU(2) Wigner functions in the $F=4$ (upper diagonal) and $F=3$ (lower diagonal) irreducible subspaces. The radius of these spheres, ranging from zero to one, determines the total population in that subspace. The off-diagonal spheres represent the coherences between the two subspaces (see text).}
\label{F:plots2}
\end{center}
\end{figure*}

At this point we have shown that we have full controllability over both the $F_+$ and the $F_-$ subspaces, as well as the 2-dimensional subspace coupled by the resonant microwaves.  We can write this in matrix form   
\be
\setlength{\unitlength}{0.25cm}   
\left(
\begin{array}{c}
\begin{picture}(16,16)(0,0)
\put(0,16){\line(1,0){9}}
\put(0,16){\line(0,-1){9}}
\put(9,16){\line(0,-1){9}}
\put(0,7){\line(1,0){9}}
\put(9,7){\line(1,0){7}}
\put(9,7){\line(0,-1){7}}
\put(16,7){\line(0,-1){7}}
\put(9,0){\line(1,0){7}}
\put(8,8){\line(1,0){2}}
\put(8,8){\line(0,-1){2}}
\put(10,8){\line(0,-1){2}}
\put(8,6){\line(1,0){2}}
\put(3.75,11.5){$s_1$}
\put(11.75,3.25){$s_2$}
\put(10.5, 9){$\sigma$'s}
\end{picture}
\end{array} 
\right),\label{e:block}
\ee
where we have ordered the basis vectors so that the states coupled by the microwaves are adjacent to each other.  We have shown that we can simulate any operator that only has matrix elements within the three boxes in Eq.~\ref{e:block}.  This includes all diagonal operators on the whole matrix as well as all operators with only nonzero matrix elements on the super-diagonals (one off the diagonal).  The irreducible representations of angular momentum, $J_x$ and $J_y$, on the entire space have support only on the super-diagonals.  Therefore, we can simulate $J_x$ and $J_y$.  According to theorem \ref{t:rank_2} all we need to show for controllability is that we can simulate some operator with a nonzero overlap with some rank 2 operator.  Since we can simulate any diagonal operator, we can simulate $T^{(2)}_0$.  It thus follows that $\{F^{(+)}_x,F^{(+)}_y,F^{(-)}_x,F^{(-)}_y, \sigma_x, \sigma_y \}$ generates $\mathfrak{su}(d)$.

\section{Generalized Wigner function representation}

In dealing with high dimensional spin systems, it is useful to be able generate graphical representations of the quantum states which give some geometric intuition.  The spin coherent state Wigner function representation introduced by Agarwal \cite{Argarwal81} provides a generalization of the standard Wigner function based on harmonic oscillator coherent states used to describe infinite dimensional systems.  Given a spin $J$, the spin coherent state Wigner function is essentially a multipole representation on the sphere defined as,
  \be
  W_{\hat{\rho}}(\theta, \phi) = \sum_k  \sum_m \textrm{Tr} [\hat{\rho} \hat{T}^{(k)}_q(J)] Y^{(k)}_q(\theta, \phi),
  \ee 
  where $Y^{(k)}_q(\theta, \phi)$ are the spherical harmonics, and   $\hat{T}^{(k)}_q(J)$ are the irreducible spherical tensors given in Eq.~(\ref{irreducible}). For a given spin, the indices describing non-trivial irreducible tensors run from $0 \leq k \leq 2J$ and $-k \leq q \leq k$.  These plots are useful visualization tools because they capture the effect of geometric rotations on the quantum state.  Two quantum states that differ solely by a SU(2) rotation will generate Wigner functions that also differ from each other by the same physical rotation. 

We seek to generalize this to the case of a tensor product space of two spins (here electron and nuclear), equivalent to the direct sum of two irreducible representations of SU(2) in the hyperfine subspaces, $F$ and $F'$.  We achieve this by considering the expanded set of tensors defined by
  \ben
 &&T^{(k)}_{q}(F,F') = \nonumber\\ 
 &&\sqrt{\frac{2k+1}{2F+1}}\sum_{m} \braket{F,m+q}{k,q;F',m} \ket{F,m+q}\bra{F',m}.\nonumber\\
  \een 
The range of the indices is now $|F-F'| \leq k \leq F+F'$ and  $-k \leq q \leq k$.  One can easily show that for two spin manifolds, the set of operators $\{ T^{(k)}_{q}(F,F), T^{(k)}_{q}(F,F'), T^{(k)}_{q}(F',F), T^{(k)}_{q}(F',F')\}$ comprises a complete orthonormal operator basis for the tensor product space.  We again can map these operators to the spherical harmonics, and for each state get four spherical Wigner functions: one each for the $F$ and $F'$ manifolds, and two for the coherences between manifolds.  We label them $W_{F,F},W_{F,F'},W_{F',F}$ and $W_{F',F'}$.  By the Hemiticity of the density operator, $W_{F,F'}$ and $W_{F',F}$ contain redundant information and are complex, so one need only consider the real and imaginary part of $W_{F,F'}$, yielding four real functions.  

We scale the radii of the spheres over which the Wigner function is plotted.  For the functions that describe a given hyperfine manifold, we let the radius of the sphere equal the population in the subspace, $\textrm{Tr}(P_F \rho P_F)$. In order to set the radii of the spheres corresponding to the coherences between the manifolds, we look at the sum of the singular values of the off-block component of the density matrix, $\sqrt{\sum_m \sum_{m'} |\bra{F,m} \rho \ket{F',m'}|^2}$.  This allows for nonequal dimensions of the two subspace.  Additionally, we scale these ``coherence spheres" by the ratio of the real versus imaginary parts of Wigner function.  The primary purpose of doing this is to be able to distinguish between pure superpositions and incoherent mixtures between the two manifolds.

To gain some intuition, we show examples of different states and different representations.  Figure~(\ref{F:plots1}) shows bar charts of the absolute values of the density matrix elements for the six states: $\ket{\psi}_{ai} = \ket{4,4}$ and $\ket{\psi}_{aii} = (\ket{4,4}+ \ket{3,-3})\sqrt{2}$ are spin coherent states and their superposition;~ $\ket{\psi}_{bi}$, and $\ket{\psi}_{bii}$ are superpositions of spin squeezed states in the two manifolds along different quadratures;~  $\ket{\psi}_{ci}$, and $\ket{\psi}_{cii}$ are coherent superpositions vs. incoherent mixtures of a ``cat state"  $(\ket{3,3}+\ket{3,-3})/\sqrt{2}$ in one manifold, and a Dicke state $\ket{4,0}$ in the other.  The corresponding Wigner functions are shown in Fig.~(\ref{F:plots2}).  From these plots we make the following observations.  When restricted to a subspace corresponding to a given hyperfine manifold, the Wigner functions on the diagonal have the familiar forms of $SU(2)$ Wigner functions, with the radius of the sphere determining the total population in that subspace.  The off-diagonal Wigner functions show the effect of the coherences, had the entire Hilbert space been determined by an irreducible representation.  This is clearly seen in Fig.~(\ref{F:plots2}aii), where the coherences are of the familiar form for a superposition of ``north" and ``south" pole spin coherent states.  The effect of geometric rotation is exhibited in $\ket{\psi}_{bi}$ and $\ket{\psi}_{bii}$.  The bar charts do not indicate any similarity between the states, while the Wigner functions are clearly related by a 90 degree rotation.  Finally, the difference between coherent superpositions and incoherent mixtures of states in the two manifolds is clearly seen in  Fig.~(\ref{F:plots2}c).
 
\bibliography{mwrf}

\end{document}